\documentclass[aps.pra,twocolumn,showpacs,preprintnumbers,amsmath,amssymb]{revtex4}
\usepackage{mathrsfs}
\usepackage{graphicx}
\usepackage{amsmath}
\usepackage{ulem}
\usepackage{color}

\begin{document}

\title{Superconducting qubit in a nonstationary transmission line cavity: parametric excitation, periodic pumping, and energy dissipation}
\author{A. A. Zhukov$^{1,2}$, D. S. Shapiro$^{1,3,4,5}$, S. V. Remizov$^{1,3}$, W. V. Pogosov$^{1,4,6}$, Yu. E. Lozovik$^{1,2,4,7}$}
\affiliation{$^1$N. L. Dukhov All-Russia Research Institute of Automatics, 127055 Moscow, Russia}
\affiliation{$^2$National Research Nuclear University (MEPhI), 115409 Moscow, Russia}
\affiliation{$^3$V. A. Kotel'nikov Institute of Radio Engineering and Electronics, Russian Academy of Sciences, 125009 Moscow, Russia}
\affiliation{$^{4}$Moscow Institute of Physics and Technology, Dolgoprudny, Moscow Region 141700, Russia}
\affiliation{$^{5}$National University of Science and Technology MISIS, 119049 Moscow, Russia}
\affiliation{$^6$Institute for Theoretical and Applied Electrodynamics, Russian Academy of
Sciences, 125412 Moscow, Russia}
\affiliation{$^7$Institute of Spectroscopy, Russian Academy of Sciences, 142190 Moscow region,
Troitsk, Russia}

\begin{abstract}
We consider a superconducting qubit coupled to the nonstationary transmission line cavity with modulated frequency taking into account energy dissipation.
Previously, it was demonstrated that in the case of a single nonadiabatical modulation of a cavity frequency
there are two channels of a two-level system excitation which are due to the absorption of Casimir photons
and due to the counterrotating wave processes responsible for the dynamical Lamb effect.
We show that the parametric periodical modulation of the resonator frequency can increase dramatically
the excitation probability. Remarkably, counterrotating wave processes under such a modulation start to play an important role even in the resonant regime.
Our predictions can be used to control qubit-resonator quantum states as well as to study experimentally different channels of a parametric qubit excitation.
\end{abstract}

\author{}
\maketitle
\date{\today }

\section{Introduction}

Quantum electrodynamics (QED) of superconducting circuits is one of fast and intensively developing fields of a modern physics.
The interest to superconducting circuits, which consist of Josephson qubits and transmission line cavities \cite{MSS},
is heated by the possibility of implementation of quantum computation \cite{Martinis},
observation of new phenomena of quantum optics in GHz frequency domain \cite{Oelsner},
as well as an engineering of sub-wavelength quantum metamaterials \cite{Macha}.
An outstanding feature of superconducting circuits
is that their parameters  are tunable {\it in situ}: excitation frequencies of qubits can be
varied externally, while both the frequency of fundamental mode of a resonator and qubit-resonator coupling energy can be modulated in GHz range by means of auxiliary SQUIDs embedded in the circuit's architecture or using more sophisticated methods. Particularly, superconducting quantum circuits can be used as a unique platform to investigate nonstationary cavity QED phenomena, such as the dynamical Casimir effect \cite{DCE2}.

In the series of papers \cite{Pokrovski1, Lozovik1} dealing with optical systems there was considered a behavior of a two-level atom in a nonstationary high-Q cavity, which experiences a single nonadiabatic change of its frequency. One of the channels of a parametric atom excitation in this situation is through a nonadiabatic change of its Lamb shift, which was termed the "dynamical Lamb effect" \cite{Lozovik1}. It is produced by counterrotating wave processes leading to a modulation of the atom's dressing by virtual photons and can be considered as the new effect in the nonstationary cavity QED.
There is another mechanism of atom excitation in this system which is
due to the absorption of photons generated by the cavity dynamical Casimir effect \cite{Lozovik1}.
The absorption is governed by resonant (Jaynes-Cummings) processes.
This mechanism is, in general, dominant for the case of nonstationary cavity and therefore it "screens" the dynamical Lamb effect.

In our recent papers \cite{paper1} (see also Ref. \cite{Berman}), we suggested an idea how to make the dynamical Lamb effect dominant. It is attractive to use a superconducting system which consists of a {\it stationary} resonator having a tunable coupling with the qubit. No Casimir photons are generated in this case, while the only one channel of qubit excitation is through the dynamical Lamb effect. Although a proposed idea allows for the observation of this effect, its unambiguous experimental realization may be not so easy.

Therefore, it is of interest to come back to a simpler scheme with variable resonator frequency, which is more straightforward to implement. In this article, we concentrate on the effect of a {\it periodic} modulation of the cavity mode frequency. We show that it provides a tool to distinguish between different channels of qubit excitation even near the resonance as well as to enhance the effect as a whole. We also take into account both energy dissipation and pure dephasing, which always exist in real systems and are able to suppress quantum effects. In contrast to most of other studies, we mainly focus on the analysis of different mechanisms of a parametric qubit excitation, i.e., due to rotating wave processes and counterrotating wave processes and under the variation of only the resonator frequency.

\section{System}

The effect under consideration can be implemented in tunable superconducting circuits, see, e.g., Ref. \cite{Nori}. As it is shown in Fig. 1 (a,b), the basic components of possible setups involve single mode cavity (superconducting coplanar waveguide), which has auxiliary SQUID embedded into one of its ends, and an artificial macroscopic atom, such as flux qubit (a) or transmon (b), coupled inductively or capacitively to the cavity. Equivalent electric circuit of the resonator is associated with $LC$-contour inside the red dashed sector in Fig 1. Alternating external flux $\Phi(t)$, threading the SQUID loop, provides an effective modulation of the resonator inductance at the desired frequency. As a consequence, such a modulation via SQUID plays a role of a non-stationary boundary conditions for the electromagnetic field in the cavity. Eventually, this results in modulation of the photon mode frequency.

\begin{figure}[h]
\center\includegraphics[width=0.95\linewidth]{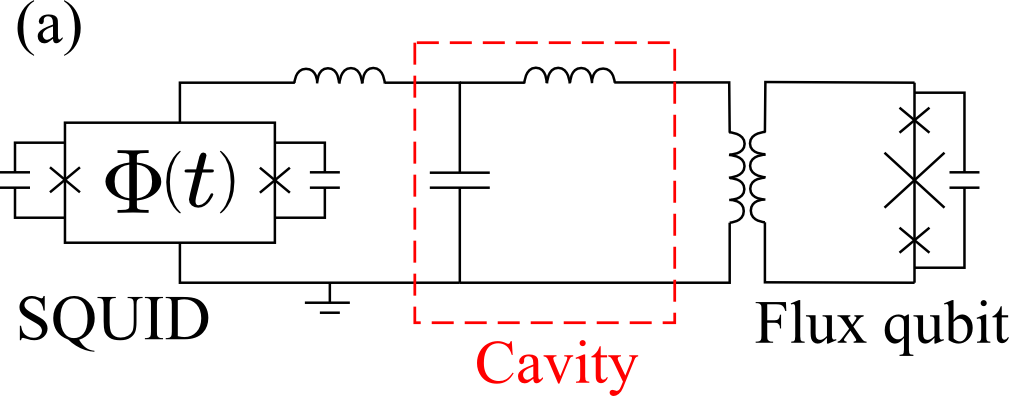}
\center\includegraphics[width=0.95\linewidth]{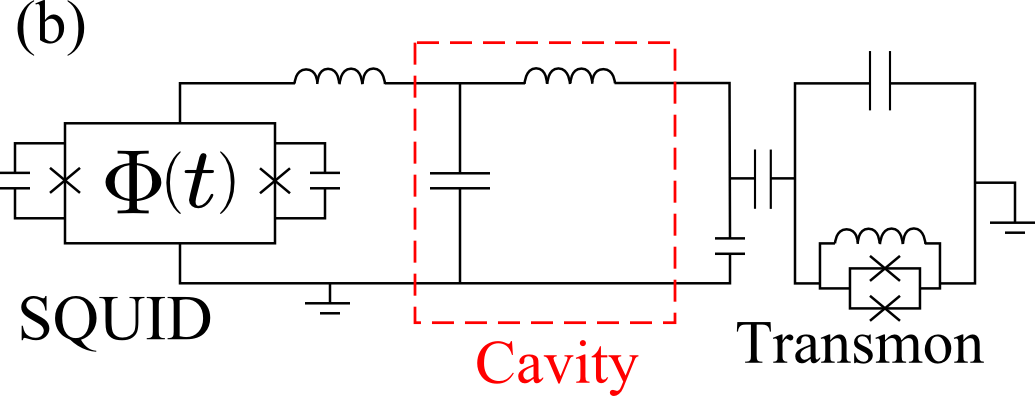}
\caption{
\label{scheme}Equivalent electric superconducting circuits of the systems under consideration. Both setups (a,b) consist of: 1) auxiliary SQUID, 2) single mode cavity, represented as $LC$-contour (red dashed), and 3) flux qubit (a) or transmon (b). Crosses stand for Josephson junctions with the sizes being related to the values of Josephson energies. SQUID's loop is subjected to the rapidly tunable magnetic flux $\Phi(t)$. The interaction between the electromagnetic field in the cavity and qubits can be organized via (a) inductive coupling in case of the flux qubit or (b) via capacitive coupling in case of the transmon.
}
\end{figure}

\section{Model}

The full non-stationary Hamiltonian of the system under consideration can be represented as
\begin{equation}
H(t)=H_0(t)+H_{\mathrm{Cas}}(t)+V.
\label{Hamiltonian}
\end{equation}
The Hamiltonian of non-interacting qubit and cavity is given by
\begin{equation}
H_0(t)=\hbar\omega(t) a^{\dagger }a + \frac{1}{2} \epsilon (1+\sigma_{3}),
\end{equation}
where $a^{\dagger }$ and $a$ are secondary quantized operators of  photon creation and annihilation in the transmission line cavity of non-stationary frequency $\omega(t)$.
 Pauli operators $\sigma_{3}=2\sigma_{+}\sigma_{-}-1$, $\sigma_{+}$, $\sigma_{-}$ act in the space of qubit excited and ground states.
The non-stationary term $H_{\texttt{Cas}}(t)$ in (\ref{Hamiltonian}) is responsible for the dynamical Casimir effect, i.e., the photon generation from vacuum  \cite{Law, Lozovik-plasma, VDodonov, ADodonov}
\begin{equation}
H_{\texttt{Cas}}(t)=i\hbar\frac{\partial_t\omega(t)}{4\omega(t)}(a^2-a^{+2}).
\end{equation}
The last term $V$ in (\ref{Hamiltonian}) describes a qubit-cavity interaction
\begin{equation}
V=g(a+a^{\dagger })(\sigma_{-}+\sigma_{+}),
\label{FullV}
\end{equation}
where $(a+a^{\dagger })$ and $(\sigma_{-}+\sigma_{+})$ can be associated with the electric field and dipole
moment, respectively, while $g$ is the coupling energy. This
interaction term can be divided into two parts $V=V_1+V_2$,
where $V_1= g(a \sigma_{+} + a^{\dagger } \sigma_{-})$
yields the well known rotating wave approximation (RWA) or Jaynes-Cummings model, provided $V_2$ is dropped, while $V_2$ is given by $V_2=g(a^{\dagger } \sigma_{+} + a \sigma_{-})$. RWA terms conserve the total excitations number, whereas counterrotating wave contributions produce and annihilate pairs of excitations.

As it was shown in \cite{Lozovik1}, in the case of a single instantaneous switching of cavity frequency $\omega$ from $\omega_1$ to $\omega_2$, the qubit excitation probability at $t \rightarrow \infty$ due to the Jaynes-Cummings processes (absorption of Casimir photons generated by $H_{\texttt{Cas}}(t)$) strongly depends on $\omega_2$ as
\begin{equation}
w^{(C)}_{e} \simeq \frac{g^2}{(\varepsilon-\omega_2)^2}\frac{(\omega_2-\omega_1)^2}{4\omega_1\omega_2},
\label{wC}
\end{equation}
when $|\varepsilon-\omega_2| \gg g$. It turns out that in the opposite case $|\varepsilon-\omega_2| \ll g$ the maximum value  $w^{(C)}_{e} \sim (\omega_2-\omega_1)^2 / \omega_{2}^{2}$ is achieved in the resonance between $\varepsilon$ and $\omega_2$ \cite{Lozovik1}. Note that this last value is independent on $g$ and, in the case of a weak modulation is small.

The qubit excitation probability due to the counterrotating wave processes, i.e., the dynamical Lamb effect is not so strongly dependent on $\omega_2$ \cite{Lozovik1}:
\begin{equation}
w^{(L)}_{e} \simeq g^2\ \frac{(\omega_2-\omega_1)^2}{(\omega_2+\varepsilon)^2(\omega_1+\varepsilon)^2},
\label{wL}
\end{equation}
which in principle allows for the separation of the two effects: $w^{(L)}_{e}$ becomes of the order of $w^{(C)}_{e}$ at large detuning $|\varepsilon-\omega_2| \sim \omega_2$. But $w^{(L)}_{e}$ is small as $\sim (\omega_2-\omega_1)^2 g^2 / \omega_{2}^{4} $. At $g/\omega_2 \ll 1$, this quantity is much smaller than the maximum value of $w^{(C)}_{e}$ attained near the resonance, where the excitation probability is controlled by Jaynes-Cummings processes. These circumstances make it problematic to probe the mechanism of qubit excitation linked to counterrotating terms.

Now we consider a {\it periodic} modulation of resonator frequency
\begin{equation}
\omega(t) = \omega_{0} + d \cos(\Omega t).
\label{omega}
\end{equation}
 There appear several controlling parameters: the time-averaged detuning $\Delta = \varepsilon - \omega_{0}$, modulation frequency $\Omega$, and its amplitude $d$. We hereafter concentrate on the limits of a small-amplitude variations, $d \ll \omega_{0}$, and a weak qubit-cavity coupling, $g \ll \omega_{0}$. We then address system's dynamics by solving numerically the Lindblad equation
\begin{equation}
\partial_t\rho(t)-\Gamma[\rho(t)]=-i[H(t),\rho(t)],
\label{Lindblad}
\end{equation}
where $\rho(t)$ is a density matrix of qubit and photon mode. Dissipation in the cavity of the rate $\kappa$ and qubit decoherence $\gamma$ are described through the matrix $
\Gamma[\rho]=\kappa(a\rho a^{\dagger } -\{ a^{\dagger }a, \rho\}/2)+\gamma(\sigma_-\rho\sigma_+ - \{\sigma_+\sigma_-,\rho\}/2)+ \gamma_{\varphi}(\sigma_z\rho\sigma_z-\rho)$.
In superconducting qubits the pure decoherence rate $\gamma_{\varphi}$ is typically
of the same order as relaxation $\gamma$. Both quantities are significantly larger than the relaxation rate of a cavity, $\gamma\gg \kappa$.

\section{Results}

If a parameter of modulation $d$ is small enough, our calculations show that both the mean number of photons $n_{ph}(t)$ and qubit excited state occupation $w_{e}(t)$ tend to constant values as $t \rightarrow \infty$. Such a situation is illustrated in Fig. 2, where we plot $\Delta w_{e}(t)=w_{e}(t)-w_{e}(0)$ and $\Delta n_{ph}(t)=n_{ph}(t)-n_{ph}(0)$, while $t=0$ corresponds to the beginning of a modulation; $T_{R}=\pi/g$ is the time scale associated with the Rabi frequency. Note that, for illustration purposes, we hereafter take into account larger $\gamma$, $\kappa$, and $\gamma_{\varphi}$ than in the state-of-art systems in order to shorten the transition time to a final regime; this does not alter a qualitative picture. A stabilization occurs due to the energy dissipation. At larger $d$, we see a change of the behavior, since no stabilization of $n_{ph}$ is achieved and external pumping overcomes total dissipation. The same behavior takes place for a resonator with modulated frequency without a qubit but having nonzero cavity relaxation rate $ \kappa$. In this case, it is not difficult to obtain explicitly the critical value of $d$ as
\begin{equation}
d_{\mathrm{crit}}^{\mathrm{(res)}} \simeq \frac{2\omega_0}{\Omega}\sqrt{\kappa^2+(\Omega-2\omega_0)^2}.
\label{dcrit}
\end{equation}
This result can be derived from Lindblad equation by reducing it to a set of equations for the number of photons $n_{ph}(t)={\rm Tr}[a^+a\rho(t)]$ and the parameter responsible for fluctuations of a photon field $a^2(t)={\rm Tr}[a^2\rho(t)]$ and then using the solution for zero frequency part $\langle n_{ph}(t)\rangle_t$, well justified in the limit $\omega_0,\omega\gg d$. Similar result has been obtained in Ref. \cite{X1}. When qubit is present in the system, the total critical value $d_{\mathrm{crit}}$ is enhanced because of the additional dissipation in the qubit, but the estimate $d_{\mathrm{crit}} \sim d_{\mathrm{crit}}^{\mathrm{(res)}}$ is still valid at $g \ll \omega_0$ and $\gamma$ not too large, as our numerical results show.

\begin{figure}[h]
 \center\includegraphics[width=0.95\linewidth]{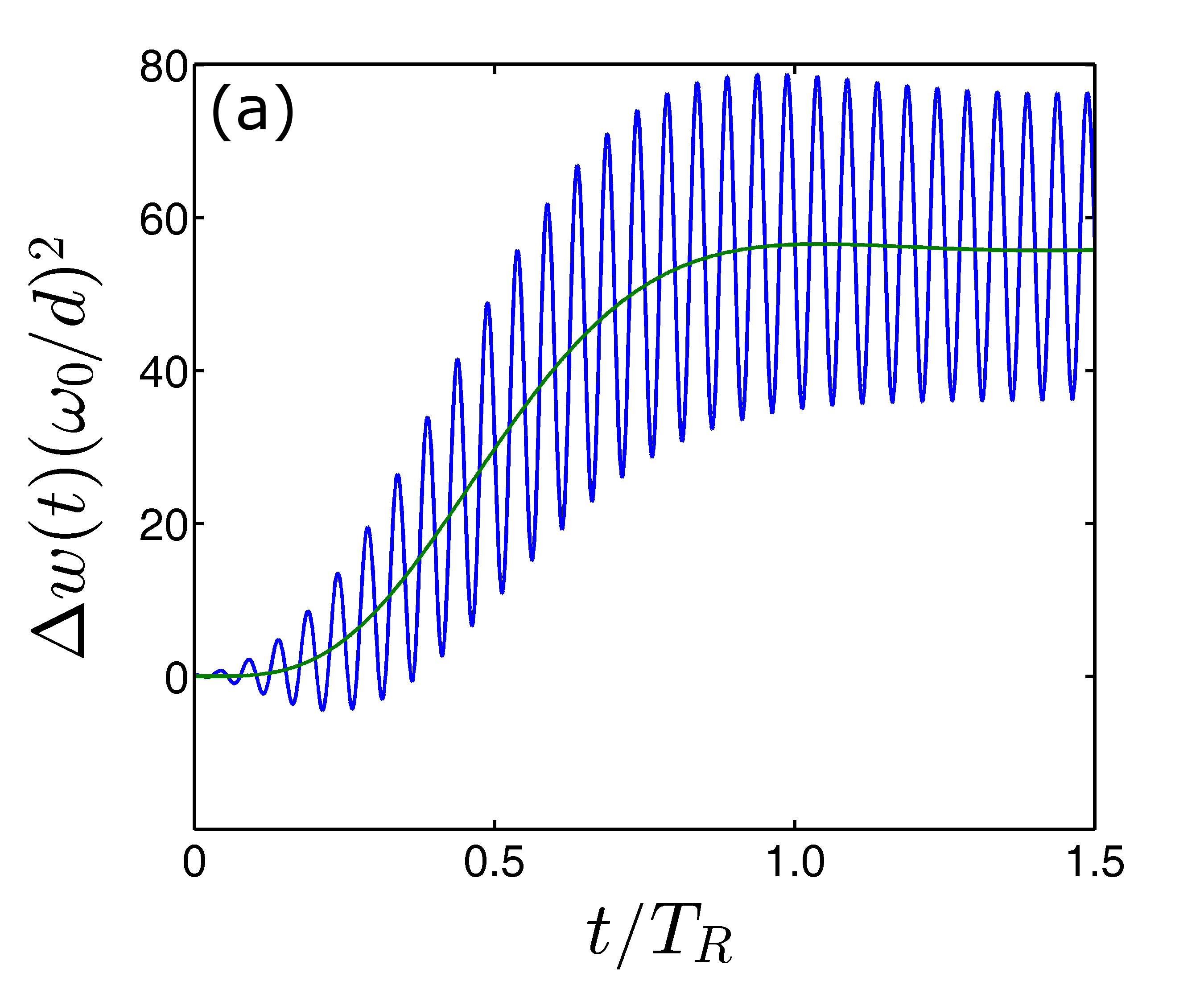}
\center\includegraphics[width=0.95\linewidth]{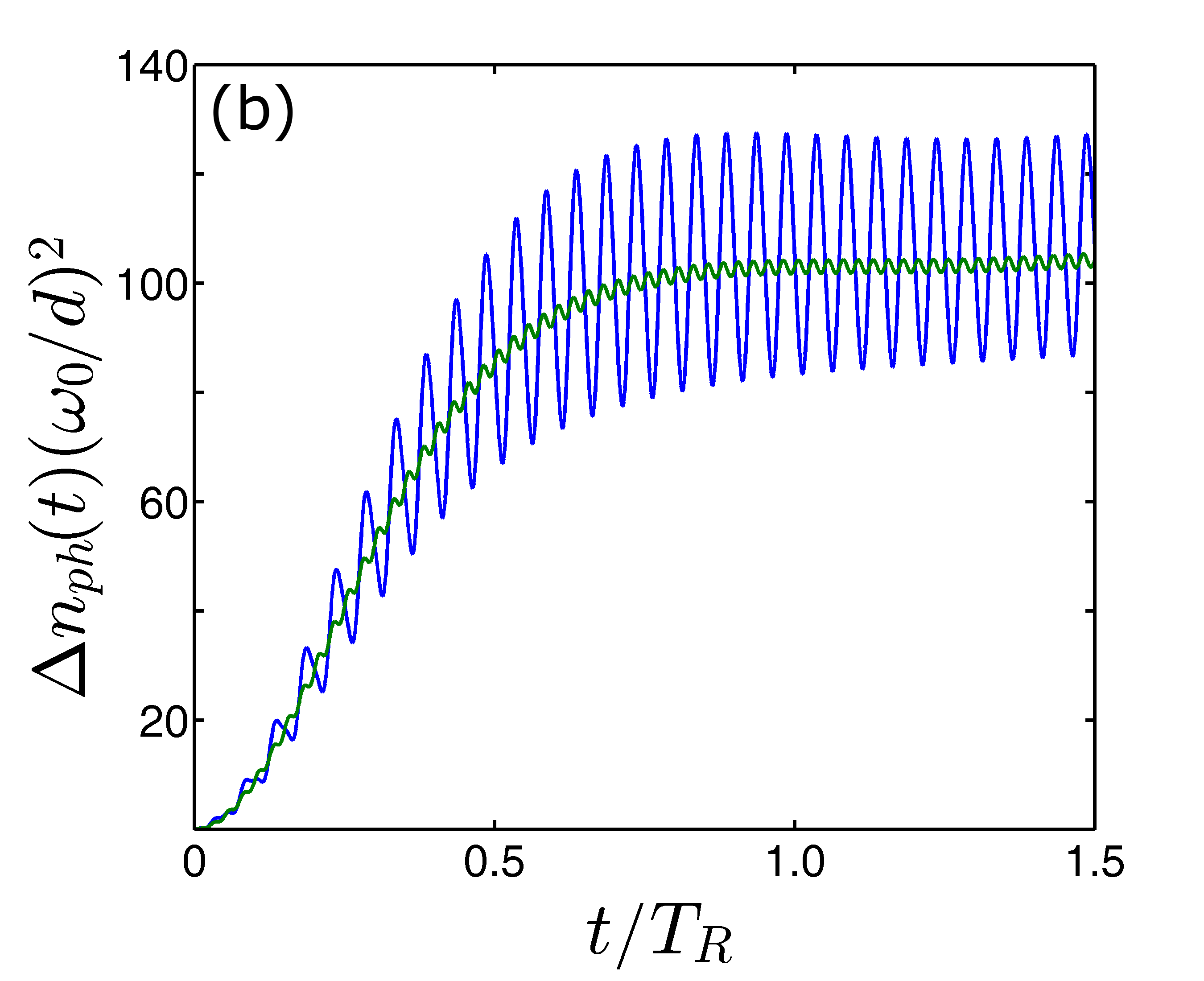} \caption{
\label{excited}(Color online) The evolution of the qubit excited state population (a)
and the mean photon number (b) after the
parametric modulation of a resonator frequency is turned on at $d=0.01\omega_{0} < d_{crit}$, $g=0.05\omega_{0}$,
$\gamma=\gamma_{\varphi}=0.05\omega_{0}$,
$\kappa=0.01\omega_{0}$,
$\Omega=2\omega_{0}$,
$\varepsilon=\omega_{0}$.
Blue highly-oscillating lines correspond to the dynamics described by the full Hamiltonian,
while green smooth lines provide similar quantities for
the Hamiltonian with only resonant (Jaynes-Cummings) terms.
}
\end{figure}

Now let us again focus on a coupled resonator-qubit system. It is reasonable to begin our considerations with the stationary limit $d=0$ and in absence of dissipation. In this case, a qubit-resonator static system can be characterized by a set of "dressed" energy levels.
It is expected that when $d$ is small but nonzero, these levels can strongly influence system dynamics provided $\Omega$
is approaching at least some of them. We indeed see in our simulations that this is the case.
For instance, if a detuning $\Delta$ is small, $w_{e}(t)$ in the final (stabilized) regime at $t \rightarrow \infty$ is mainly determined by the interplay between the dynamical Casimir effect and the Jaynes-Cummings resonant processes. In the case of $d=0$, the state $|2,g\rangle$ hybridizes with $|1,e\rangle$ mostly via RWA terms in the Hamiltonian thus forming two energy levels. Their splitting at zero detuning is determined solely by $g$, while the energy levels are positioned in the vicinity of $2 \omega_{0}$. Dynamical Casimir effect at nonzero $d$ leads to a finite occupation of the state $|2,g\rangle$. When changing $\Omega$ near $2 \omega_{0}$ we see appearing two peaks in $w_{e}(t \rightarrow \infty)$, as shown in Fig. 3 (a). These two peaks nearly correspond to the Jaynes-Cummings energy levels with two excitations. Thus, the highest $w_{e}(t \rightarrow \infty)$ can be achieved when $\Omega$ is in a resonance with one of such energy levels. This feature can be used in experiments to maximize the effect. By tuning parameters in our numerical solution, we found that this maximum value scales as $\sim \frac{d^2}{\omega_{0}^2} \frac{\omega_{0}}{\Gamma}$, where $\Gamma$ is an effective dissipation rate given by some combination of $\kappa$ and $\gamma$. If we compare this result with the result for a single switching, we see that the periodic driving increases an effect by a large factor of $\sim \omega_{0} / \Gamma$. Indeed $\omega_{2}-\omega_{1}$ is analogous to $d$, since it also represents an amplitude of a resonator frequency modulation. Hence, the large factor $\omega_{0} / \Gamma$ can be treated as a characteristic number of attempts to excite the qubit until the system is stabilized by dissipation.

 It is of interest that the widths of the two peaks as functions of $\Omega$ can be very different provided $\gamma \gg \kappa$, as usually applies for superconducting quantum circuits.
If $\Delta < 0$, the higher-energy level is mostly associated with photon degrees of freedom. However, the lower-energy state has a significant
 contribution from qubit degrees of freedom. The sub-leading contributions in both cases decrease,
 as $\Delta$ decreases. This implies that the width of the lower resonance should increase,
 while the width of the higher resonance should decrease. This is exactly the behavior we see in our solution. For illustration, corresponding peaks at larger $|\Delta|$ are plotted in Fig. 3 (b).
 Note that an effect of a pure dephasing is mainly in increasing the width of the lower-energy peak.

 \begin{figure}[h]
\center\includegraphics[width=0.95\linewidth]{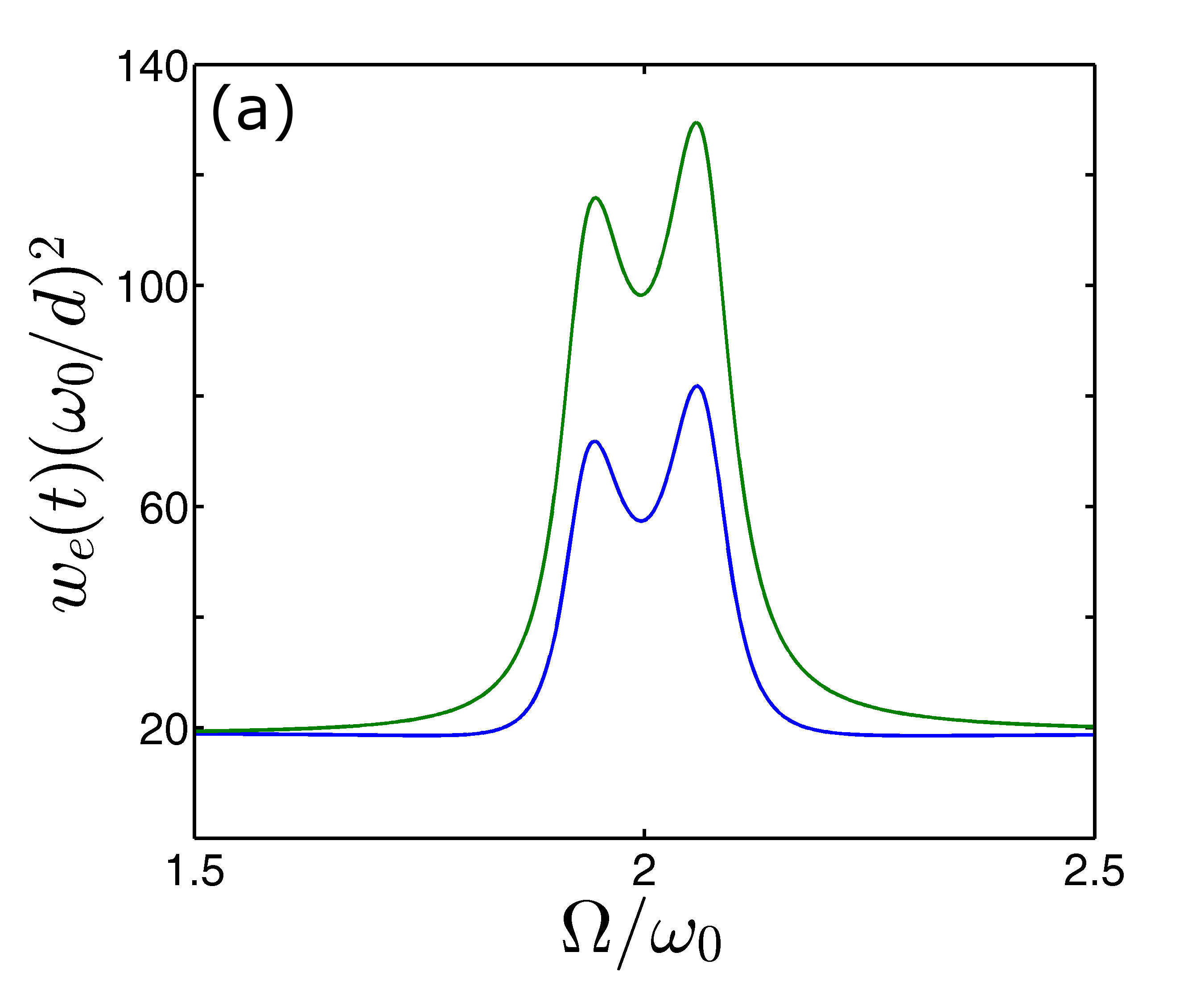}
\center\includegraphics[width=0.95\linewidth]{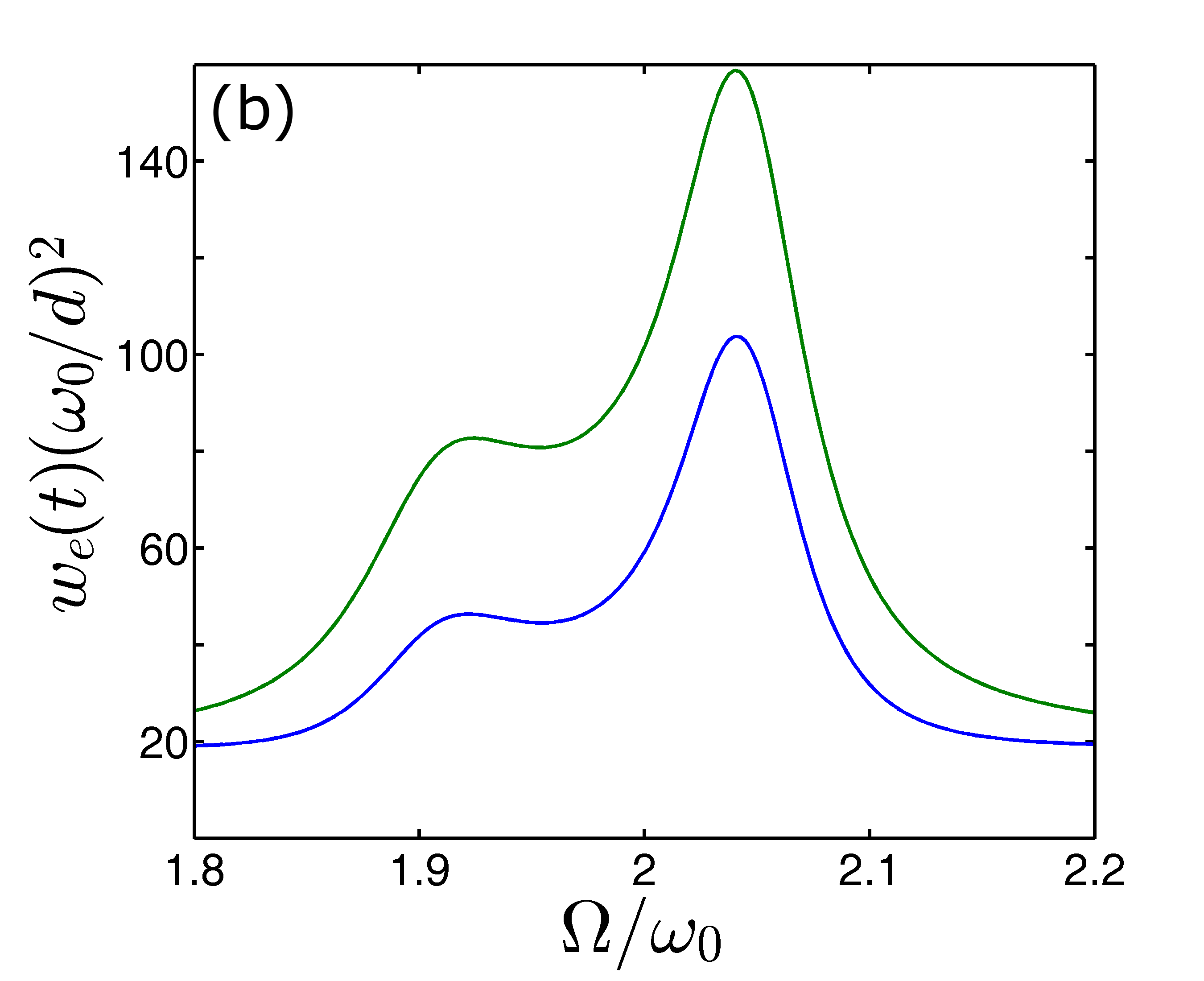} \caption{
\label{excited}(Color online) The dependence of qubit excited state population in a final regime as a function of modulation frequency at zero detuning $\Delta=0$ (a)
and $\Delta=-0.1\omega_{0}$ (b) at $g=0.05\omega_{0}$, $\gamma=\gamma_{\varphi}=0.05\omega_{0}$, $\kappa=0.01\omega_{0}$, $d=0.01\omega_{0}$. Blue upper (green lower) lines show maximum (minimum) values, between which oscillations occur.
}
\end{figure}

\begin{figure}[h]
\center\includegraphics[width=0.95\linewidth]{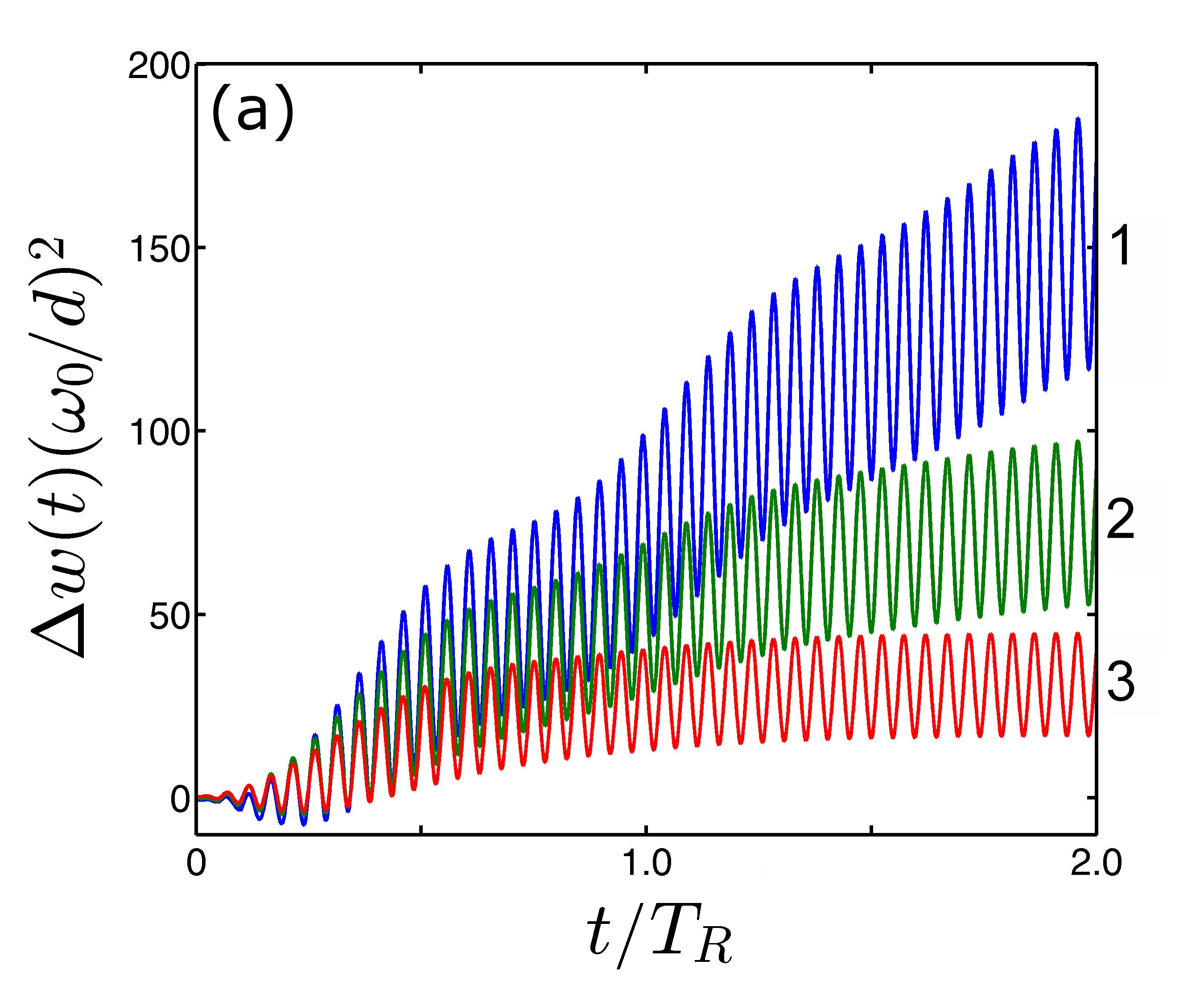}
\center\includegraphics[width=0.95\linewidth]{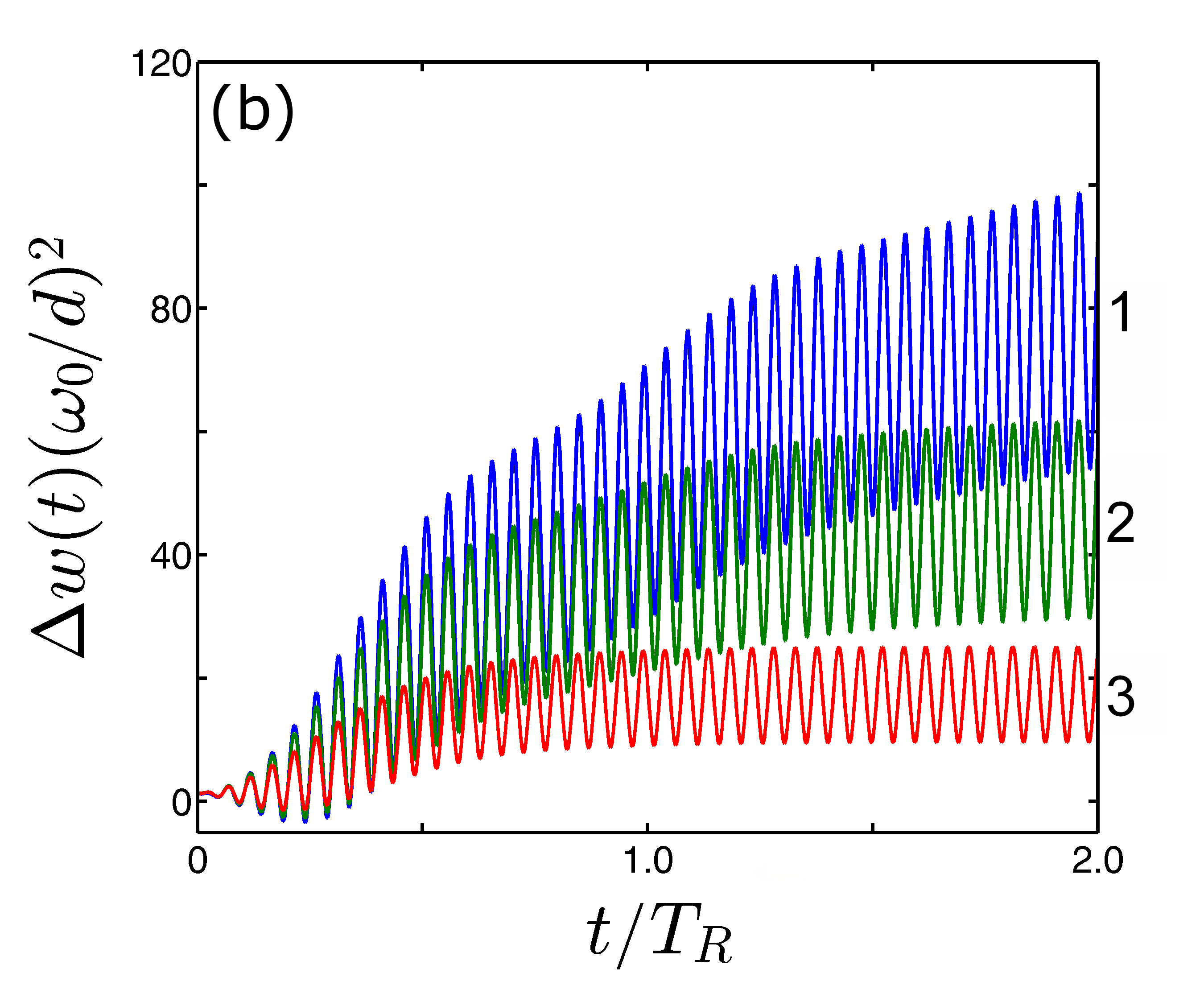} \caption{
\label{exciteddec}(Color online) The evolution of the qubit excited state population for different values of $\gamma=\gamma_{\varphi}$ at fixed $\kappa=0.01 \omega_{0}$ (a) and different values of $\kappa$ at fixed $\gamma=\gamma_{\varphi}=0.05 \omega_{0}$ (b). Blue lines 1 correspond to $\gamma=\gamma_{\varphi}=0.01 \omega_{0}$ (a) and $\kappa=0.01 \omega_{0}$ (b); green lines 2 refer to $\gamma=\gamma_{\varphi}=0.02 \omega_{0}$ (a) and $\kappa=0.02 \omega_{0}$ (b); red lines 3 stand for $\gamma=\gamma_{\varphi}=0.1 \omega_{0}$ (a) and $\kappa=0.05 \omega_{0}$ (b). In all cases, $g=0.05 \omega_{0}$, $\Omega = 2\omega_{0} + g \sqrt{2} $, $d=0.01 \omega_{0}$.}
\end{figure}

We now discuss in a more detail the influence of decoherence on system dynamics. Fig. 4 shows the time evolution of the qubit excited state population for different values of $\gamma$, $\gamma_{\varphi}$, and $\kappa$ at zero detuning $\Delta=0$ and modulation frequency $\Omega = 2\omega_{0} + g \sqrt{2} $, which corresponds to the maximum $w_{e}(t\rightarrow \infty)$, as explained before. It is clearly seen from this figure that all considered types of decoherence lead to the suppression of the maximum parametric qubit excitation. The same is true for the number of photons generated from vacuum. Notice, however, that if $\Omega$ is not in a resonance with the Jaynes-Cummings energy levels (for instance, at $\Omega = 2\omega_{0}$), decoherence can increase $w_{e}(t \rightarrow \infty)$ due to the smearing of two peaks seen in Fig. 3.

Despite the fact that RWA physics plays an important role near the resonance between $\varepsilon$ and $\omega_{0}$, counterrotating terms also lead to remarkable effects. They are responsible for fast and rather significant in their amplitude oscillations of $w_{e}$ as a function of time, although the whole dependence in general follows the trend determined by Jaynes-Cummings processes, as seen from Figs. 2, 3, and 4. The amplitude of these oscillations is nearly proportional to both $d$ and $g$. The oscillations appear also in the temporal dependence of $n_{ph}$. This result is unexpected, since counterrotating terms of the Hamiltonian usually can be ignored near the resonance and in the case of a weak coupling, $g \ll \omega_{0}$. In contrast, in our nonstationary system, there exists an amplification of these terms due to the periodic parametric modulation of $\omega$. We would like to stress that there is no effect of this kind in the case of a single switching, as Eqs. (\ref{wC}) and (\ref{wL}) evidence. We also note that the effect of counterrotating wave terms was recently analyzed in Ref. \cite{ADodonov} far from the resonance ("Anti-Jaynes-Cummings regime") and in Ref. \cite{Nori1} for the regime of a strong qubit-cavity coupling.

\begin{figure}[h]
\center\includegraphics[width=0.95\linewidth]{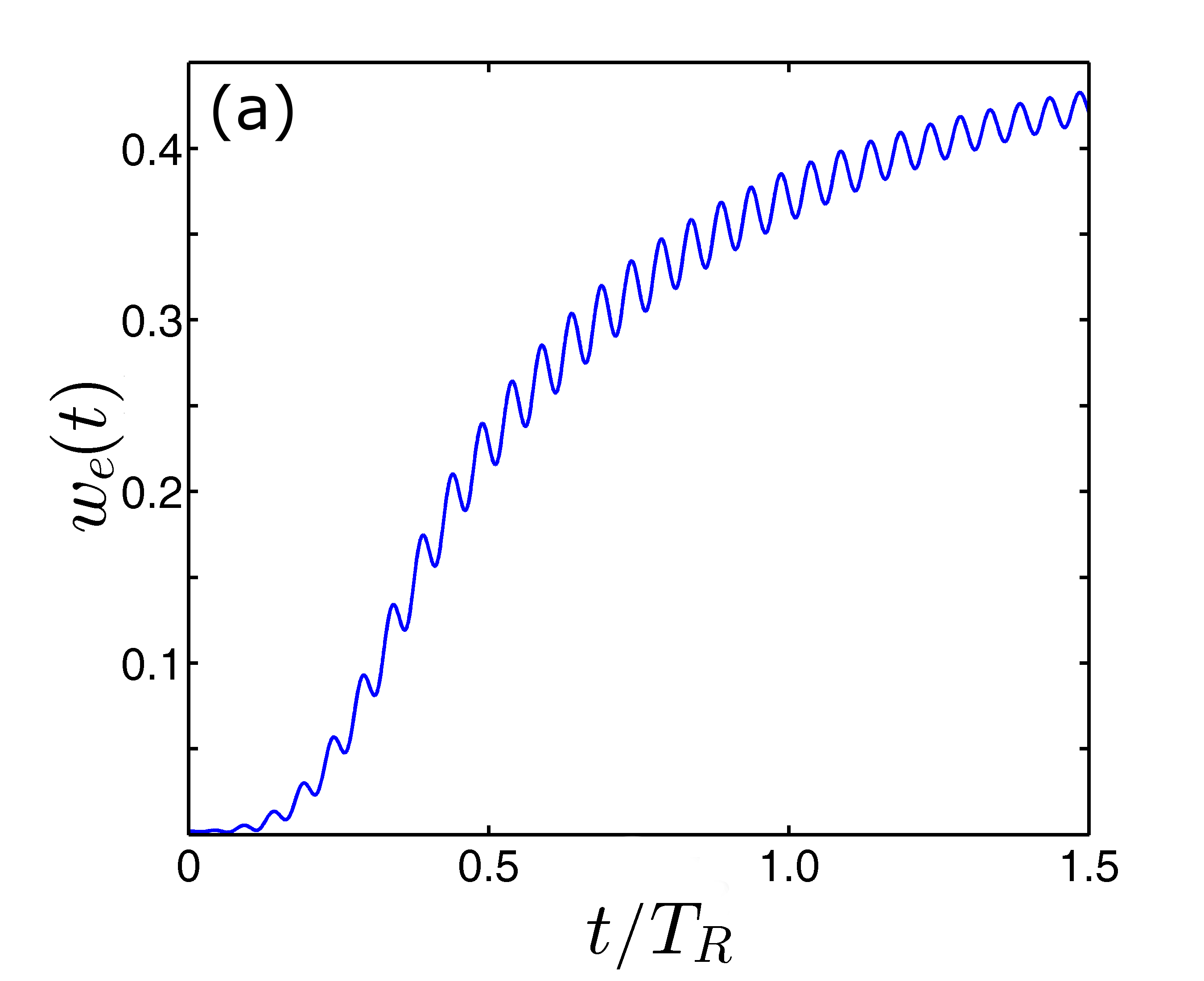}
\center\includegraphics[width=0.95\linewidth]{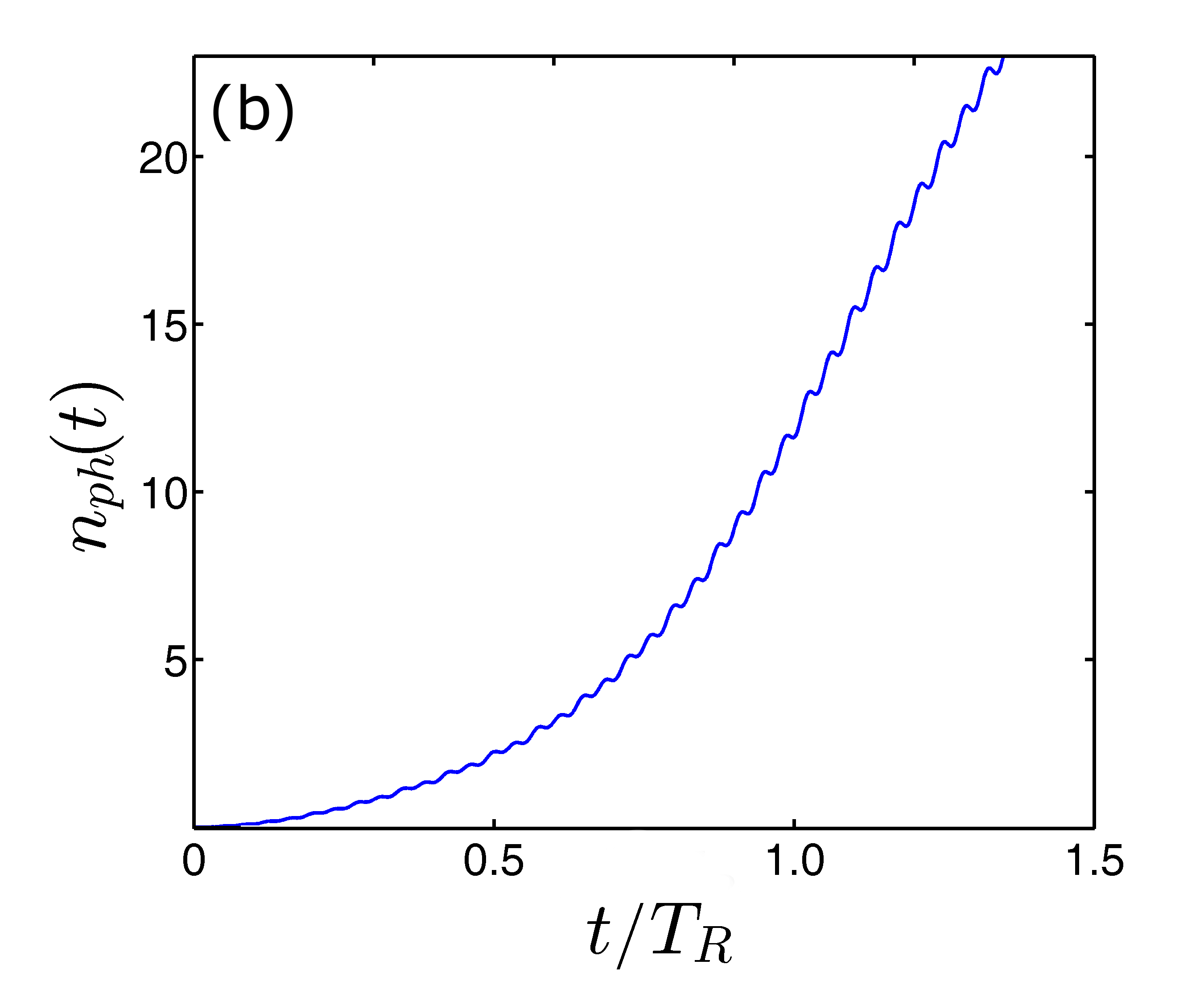} \caption{
\label{excited}(Color online) The qubit excited state population (a)
and the mean photon number (b) as functions of time after the
parametric modulation of a resonator frequency is turned on at $d=0.1\omega_{0} > d_{crit}$, $g=0.05\omega_{0}$,
$\gamma=\gamma_{\varphi}=0.05\omega_{0}$,
$\kappa=0.01\omega_{0}$,
$\Omega=2\omega_{0}$,
$\varepsilon=\omega_{0}$.
}
\end{figure}

The oscillations we found can be studied in experiments and they may serve as a tool to distinguish between the rotating wave and counterrotating wave physics in the case of a parametric driving. However, it is not easy to resolve them in experiments using spectroscopic approaches because of GHz frequencies of these oscillations.
Fortunately, there exist measurement techniques with time resolution up to 1 picosecond based on bifurcation oscillators or Josephson ballistic interferometers \cite{Siddiqi}. Such tools should allow to probe the dynamical Lamb effect in state-of-the-art or near future superconducting quantum circuits.
Similar fast oscillations exist at large detunings, where total qubit excited state occupation $w_{e}$ is much smaller than near the resonance, in accordance with the results for the single switching. We, therefore, arrive at the counter-intuitive result that the resonant regime is preferable for observation of {\it both} channels of qubit excitation in parametrically-driven circuits with $g/\omega_{0} \ll 1$.

Let us stress that regardless of the conclusion that the periodical modulation of a resonator frequency can dramatically increase $w_{e}$, it remains small in the regime of a weak modulation. Since $w_{e}$ depends crucially on the "number of attempts" to excite the qubit, it is of importance to find a way to increase this number. Fortunately, in the case of a single resonator without a qubit if driving amplitude is large enough ($d > d_{\mathrm{crit}}^{\mathrm{(res)}}$), the number of generated photons cannot be saturated by the cavity dissipation, see Eq. (\ref{dcrit}). A qubit coupled to the resonator is unable to qualitatively change this behavior, although it provides an additional channel for the energy dissipation through $\gamma$, as discussed before. This implies that, if $d$ is large enough ($d > d_{\mathrm{crit}}$), photon number does not saturate, so that the "number of attempts" somehow goes to infinity. Our numerical results for such a regime are presented in Fig. 5. We indeed find no saturation of $n_{ph}$ in this case (b), while $w_{e}$ grows up to a large value $\approx 1/2$ (a). Such a dynamics of a qubit is expectable if it is subjected in a strong driving field. The role of this field is here played by Casimir photons. The whole dependence $w_{e}$ is again controlled by the Jaynes-Cummings terms with superimposed oscillations due to counterrotating wave processes, see Fig. 5 (a). The amplitude of such oscillations is larger than in the case of a small $d$, but the relative contribution to $w_{e}$ becomes smaller.

Our findings have to be contrasted with the results for another scheme we recently suggested \cite{paper1}, in which qubit-resonator coupling constant $g$ is varying in time instead of the resonator frequency. Within this scheme, it is possible to drive a system in a resonant regime which results in a very high $w_{e}\sim 1$ solely due to counterrotating wave processes. Thus, the two schemes we consider suggest mutually complementary approaches to study different channels of a parametric qubit excitation in nonstationary coupled qubit-resonator systems.

Finally, we would like to mention that the investigation of nonadiabatic effects in superconducting quantum circuits is of interest not only from the viewpoint of a fundamental physics, but also for purposes of quantum computation and simulation. Indeed, high-speed gates can induce nonstationary QED effects linked to the undesirable generation of excitations from vacuum, which are able to affect a device performance. Moreover, nonadiabatic effects can be used in a positive way. One of the examples is a realization of nonadiabatic holonomic quantum gates based on non-Abelian geometric phases \cite{Abdulmalikov}. Thus, the control of nonadiabatic phenomena in superconducting quantum circuits is of significant importance.

\section{Summary}

We considered a dynamics of a single qubit coupled to a resonator with time-varying frequency taking into account both energy dissipation and pure dephasing. We have shown that by using a periodic modulation of a resonator frequency, one can strongly increase the probability of a parametric qubit excitation. Surprisingly, although the qubit excited state population is mostly controlled by resonant processes, counterrotating wave processes are of importance even at small detuning, since they produce considerable oscillations of this quantity. Hence, both channels of a parametric qubit excitation, i.e., due to rotating wave and counterrotating wave terms can be probed near the resonance, when the effect of a qubit excitation is highest.

The authors acknowledge useful comments by E. Il'ichev, N. V. Klenov and K. V. Shulga. D. S. S. acknowledges a support by the Fellowship of the President of Russian Federation for young scientists (fellowship no. SP-2044.2016.5) and the Russian Science Foundation (contract no. 16-12-00095). W. V. P. acknowledges a support by RFBR (project no. 15-02-02128) and by Ministry of Education and Science of the Russian Federation (grant no. 14.Y26.31.0007).


\begin{references}

\bibitem{MSS}P. D. Nation, J. R. Johansson, M. P. Blencowe, and F. Nori, Rev. Mod. Phys.  {84} (2012) 1; Yu. Makhlin, G. Sch\" on, and A. Shnirman, Rev. Mod. Phys.  {73} (2001) 357; J.Q. You and F. Nori, Nature {474} (2011) 589.

\bibitem{Martinis}R. Barends, et al., Nat. Commun.  {6} (2015) 7654.

\bibitem{Oelsner}G. Oelsner, et al., Phys. Rev. Lett  {110} (2013) 053602; O. Astafiev, et al., Science  {327} (2010) 840; L. Zhou, et al., Phys. Rev. Lett. {101} (2008) 100501.

\bibitem{Macha}A. L. Rakhmanov, A. M. Zagoskin, S. Savel'ev, and F. Nori, Phys. Rev. B {77} (2008) 144507; P. Macha, et al., Nature  Commun. { 5} (2014) 5146.

\bibitem{DCE2}C. M. Wilson, et al., Nature  {479} (2011) 376; J.R. Johansson, G. Johansson, C.M. Wilson, and F. Nori, Phys. Rev. Lett. {103} (2009) 147003.

	\bibitem{Pokrovski1}A. A. Belov, Yu. E. Lozovik, and V. L. Pokrovsky, J. Phys. B  {22} (1989) L101-L105; A. A. Belov, Yu. E. Lozovik, and V. L. Pokrovski, Sov. Phys. JETP  {96} (1989) 552; A. M. Fedotov, N. B. Narozhny, and Yu. E. Lozovik, Phys. Lett. A  {274} (2000) 213.

	\bibitem{Lozovik1}N. B. Narozhny, A. M. Fedotov, and Yu. E. Lozovik, Phys. Rev. A  {64} (2001) 053807.

\bibitem{paper1}D. S. Shapiro, A. A. Zhukov, W. V. Pogosov, and Yu. E. Lozovik, Phys. Rev. A  {91} (2015) 063814; A. A. Zhukov, D. S. Shapiro, W. V. Pogosov, and Yu. E. Lozovik, Phys. Rev. A  {93} (2016) 063845.

\bibitem{Berman}O. L. Berman, R. Ya. Kezerashvili, and Yu. E. Lozovik, Phys. Rev. A {94} (2016) 052308.

\bibitem{Nori}J. Q. Liao, et al., Phys. Rev. A {81} (2010) 042304.

\bibitem{Law} C. K. Law, Phys. Rev. Lett. { 73} (1994)  1931.

 \bibitem{Lozovik-plasma}Yu. E. Lozovik, V. G. Tsvetus, and E. A. Vinogradov, Phys. Scr. {52}  (1995) 184.

	\bibitem{VDodonov}V. V. Dodonov, Phys. Scr. {82} (2010) 038105.

	\bibitem{ADodonov}D. S. Veloso and A. V. Dodonov, J. Phys. B: At. Mol. Opt. Phys.  {48} (2015) 165503.

\bibitem{Nori1}X. Cao, J. Q. You, H. Zheng, and F. Nori, New J. Phys. {13} (2011) 073002.

\bibitem{X1}V. V. Dodonov, Phys. Rev. A {58} (1998) 4147.

	\bibitem{Siddiqi}I. Siddiqi, et al., Phys. Rev. B. { 73} (2006) 054510; T. Picot, et al., Phys. Rev. B { 78} (2008) 132508; I. I. Soloviev, et al., Phys. Rev. B { 92} (2015) 01451.

	\bibitem{Abdulmalikov}  A. A. Abdulmalikov Jr, et al., Nature { 496} (2013) 482.

\end{references}
\end{document}